%% file: paper.tex
\documentclass[sigplan,nonacm,10pt]{acmart}
\usepackage[english]{babel}
\usepackage{blindtext}
\usepackage{graphicx}
\usepackage{colortbl}
\usepackage{xcolor}
\usepackage{pifont}
\usepackage{multirow}
\usepackage{url}
\usepackage{hyperref}
\usepackage{enumitem}

\input{macros}

\begin{document}

\title[]{Foundry: Template-Based CUDA Graph Context Materialization for Fast LLM Serving Cold Start}

\author{
\rm{
    Xueshen Liu$^{\text{1},*}$ \enskip
    Yongji Wu$^{\text{2}, *,\text{\textdagger}}$\enskip
    Yuncheng Yao$^{\text{3}}$ \enskip
    Danyang Zhuo$^{\text{3}}$ \enskip
    Ion Stoica$^{\text{2}}$ \enskip} 
    Z. Morley Mao$^{\text{1}}$ \enskip
{$^{\text{1}}$\textit{University of Michigan}\enskip$^{\text{2}}$\textit{UC Berkeley}\enskip$^{\text{3}}$\textit{Duke}}}

\renewcommand{\shortauthors}{}

\begin{abstract}
Modern LLM service providers increasingly rely on autoscaling and parallelism reconfiguration to respond to rapidly changing workloads, but cold-start latency remains a major bottleneck. While recent systems have reduced model weight loading to seconds, CUDA graph capture still takes tens of seconds to minutes and often dominates startup. Unfortunately, CUDA graphs cannot be naively serialized: beyond graph topology, they are tightly coupled to execution context, including device addresses embedded in kernel arguments and kernel code lazily loaded during warmup. Existing approaches either rely on brittle kernel-specific patching or heavyweight process-level checkpoint/restore that are inflexible to dynamic parallelism switching.

We present Foundry, a template-based CUDA graph context materialization system that persists both graph topology and execution context during an offline processing stage, and reconstructs executable graphs online with negligible overhead. Foundry enforces deterministic memory layouts, automatically extracts and reloads kernel binaries required by captured graphs, and reduces online reconstruction costs through topology-based templating. For distributed serving, Foundry further enables a single-GPU offline capture to generate templates for multi-GPU deployments by patching only rank-dependent communication state. Across dense and MoE models up to 235B parameters, Foundry reduces cold-start latency by up to 99\%, cutting the initialization time of Qwen3-235B-A22B from 10 minutes to 3.9 seconds while preserving the throughput gains of CUDA graphs. Foundry is open-sourced at \url{https://github.com/foundry-org/foundry}.
\end{abstract}

\maketitle
\pagestyle{plain}
{\let\thefootnote\relax\footnote{
$^*$ Xueshen Liu and Yongji Wu contributed equally. \\
\text{\textdagger} Corresponding author: Yongji Wu <yongji.wu@berkeley.edu>}}

\input{introduction}
\input{background}
\input{overview}

\input{design}

\input{implementation}

\input{evaluation}

\input{related}
\input{conclusion}

\bibliographystyle{ACM-Reference-Format}
\bibliography{reference}

\input{appendix}
\end{document}

%% file: macros.tex
\newcommand{\sys}{\emph{Foundry}\xspace}
\newcommand{\save}{\textbf{SAVE}\xspace}
\newcommand{\load}{\textbf{LOAD}\xspace}
\newcommand{\workspace}{archive\xspace}

%% file: introduction.tex
\section{Introduction}
The rapid progress of large language models (LLMs) is reshaping a wide range of applications, from chatbots and coding assistants to browser agents. At the same time, serving LLMs at scale incurs substantial infrastructure cost—recent reports estimate that OpenAI alone has spent roughly \$12B on inference compute since 2024~\cite{zitron2025oai_docs}. This cost is exacerbated by the highly dynamic and unpredictable nature of LLM serving workloads: request rates and sequence lengths can shift dramatically over minutes~\cite{qiao2024conserve,yu2025prism,wang2025burstgpt}. As a result, static provisioning (e.g., allocating a fixed number of GPUs per model) often leads to low utilization~\cite{qiao2024conserve}. Moreover, the optimal parallelism configuration depends on workload characteristics, with different strategies trading off latency, throughput, memory footprint, and communication overhead~\cite{chen2025gyges}.

To improve resource utilization, modern LLM serving stacks increasingly rely on elasticity. Autoscaling mechanisms scale the number of serving instances in response to real-time loads~\cite{yu2025lambda,lou2026hydraserve,lou2025warmserve,fu2024serverlessllm,hu2025deepserve}. Complementarily, parallelism hot switching techniques dynamically reconfigure the parallelism strategy to better match the current workload characteristics~\cite{chen2025gyges,liu2025expert,wu2024loongserve}. Together, these approaches aim to sustain high utilization while meeting latency/throughput SLOs under rapidly changing traffic patterns.

However, both scenarios are bottlenecked by cold start latency~\cite{lou2026hydraserve,zeng2025medusa}. Launching a new serving instance—or reconfiguring an existing one—requires (i) loading (resharding) model weights and (ii) (re)capturing GPU execution graphs, e.g., CUDA graphs on NVIDIA GPUs or HIP graphs on AMD GPUs. Recent systems reduce weight loading to 1–2 seconds via RDMA-based transfer from peer instances, even for trillion-parameter models such as Kimi-K2~\cite{perplexity-rdma-weight-transfer,lmsys-rfork}. In contrast, graph capture can take several minutes and often dominates end-to-end cold start time~\cite{zeng2025medusa}.

CUDA graphs have become the de facto solution employed by inference frameworks to reduce the CPU kernel launching overhead~\cite{vllm-cuda-graphs}. With CUDA graphs, a series of independent kernel launches are grouped into a single launching unit. They significantly improve the performance of LLM decoding inference. However, CUDA graphs require a time-consuming capturing forward to construct. 
Concretely, the inference framework executes the CPU-side model forward logic, while the CUDA runtime intercepts kernel launches and records their functions, parameters, and execution dependencies to construct the graph. Because graph capture still performs extensive CPU operations as in a normal forward pass, it incurs significant latency overhead~\cite{zeng2025medusa}.

At first glance, this problem seems amenable to simple serialization: the system could dump the topology of captured CUDA graphs offline and reload them during online serving. However, a CUDA graph is not merely a topology description; it is tightly coupled to the execution context in which it was captured. In particular, graph nodes may reference device-side resources, including memory pointers embedded in kernel arguments and kernel function handles resolved by the CUDA runtime. These context-dependent references make CUDA graphs inherently non-portable and prevent straightforward serialization.

Existing approaches explore two different designs to mitigate this problem. One line of work, exemplified by Medusa~\cite{zeng2025medusa}, adopts a \textit{patch-based graph restoration} mechanism. It materializes the graph topology, but without any device-side resources. At online serving, it applies a hand-crafted rule to manually trigger the loading of relevant device resources and input parameters for each used kernel. Reliance on per-kernel patching rules makes Medusa difficult to support rapidly evolving hardware platforms and model architectures, which frequently introduce new custom kernels. At the other end of the spectrum, \textit{process-level checkpointing} supported by modern GPU drivers, e.g., NVIDIA's cuda-checkpoint~\cite{nvidia2025cudacheckpoint}, can be leveraged to snapshot a LLM serving instance after it is initialized. However, process checkpointing produces substantially larger checkpoint images as it blindly bundles all GPU and CPU states across all worker processes. It also fails at dynamic parallelism switching, as the states of in-flight requests are lost when restoring the full worker processes.

In addition, model providers typically serve many models under a wide array of parallelism strategies~\cite{mooncake2026elasticep,liu2025expert,chen2025gyges}, while they frequently integrate new kernel and engine optimizations to improve serving performance. Hence, \textit{offline processing itself must be economical.} It is infeasible to frequently spin up the full target number of GPUs for each model under each parallelism configuration, just to create checkpoint images or graph topology files.

To address these limitations, we present \sys, a system that persists CUDA graph states through \textit{template-based context materialization}. \sys materializes both the topology and execution context of the captured graphs, making graph restoration kernel-agnostic and eliminating the need for hand-crafted patching rules. To minimize the cost of offline preprocessing, \sys enables a single GPU to materialize graph \textit{templates} for distributed inference under different parallelism configurations. During online serving, \sys instantiates these templates into executable graphs for each GPU by configuring the rank-dependent state of communication libraries (e.g., NCCL~\cite{nvidia_nccl_github} or NVSHMEM~\cite{nvidia_nvshmem_github}) referenced by the graphs.

However, several challenges remain in realizing \sys. First, how can \sys materialize execution context in a general manner that supports arbitrary kernels? Second, how can \sys transparently enable a single-GPU worker process to record graph templates for multi-GPU inference? Third, how can \sys minimize online restoration overhead, given that inference engines typically capture many graphs spanning a wide range of batch sizes to optimize performance?

To address the first challenge, \sys intercepts CUDA memory allocations and redirects them to virtual memory management (VMM) APIs to enforce a deterministic memory layout; in addition, \sys automatically serializes and restores the in-memory binaries of the kernels used in the captured graphs. To address the second challenge, \sys introduces a stub layer over communication libraries to emulate distributed communication during single-GPU offline processing. Finally, \sys further reduces restoration overhead by allowing graphs with the same topology to share a common template and by applying in-place updates to graph-node parameters.

We implement \sys in PyTorch and prototype its integration with vLLM~\cite{kwon2023pagedattention} to persist CUDA graphs. We evaluate \sys using both dense and MoE models under data and expert parallelism. The results show that \sys reduces vLLM cold-start latency by up to 99\%, lowering the initialization latency of Qwen3-235B-A22B~\cite{yang2025qwen3} from 10 minutes to 3.9 seconds, while preserving the inference performance gains of CUDA graphs.

In summary, we make the following contributions:
\begin{itemize}[leftmargin=*]
    \item We propose \textit{context materialization} to mitigate CUDA graph capture overhead in LLM serving cold start. By persisting both graph topology and execution context, \sys enables graph restoration for arbitrary kernels without kernel-specific patching rules.
    
    \item We develop a \textit{template-based} graph restoration mechanism that enables a single GPU to record graph templates for multi-GPU inference. In addition, graphs captured at different batch sizes can share a template when they have the same topology, reducing restoration overhead.
    
    \item We comprehensively evaluate \sys, demonstrating its effectiveness in reducing LLM serving cold-start latency across a range of deployment scenarios.
\end{itemize}

%% file: background.tex
\section{Background and Motivation}
\label{sec:background}

\subsection{LLM Serving Cold Start}

\begin{figure}[t]
\centering
\includegraphics[width=0.99\linewidth]{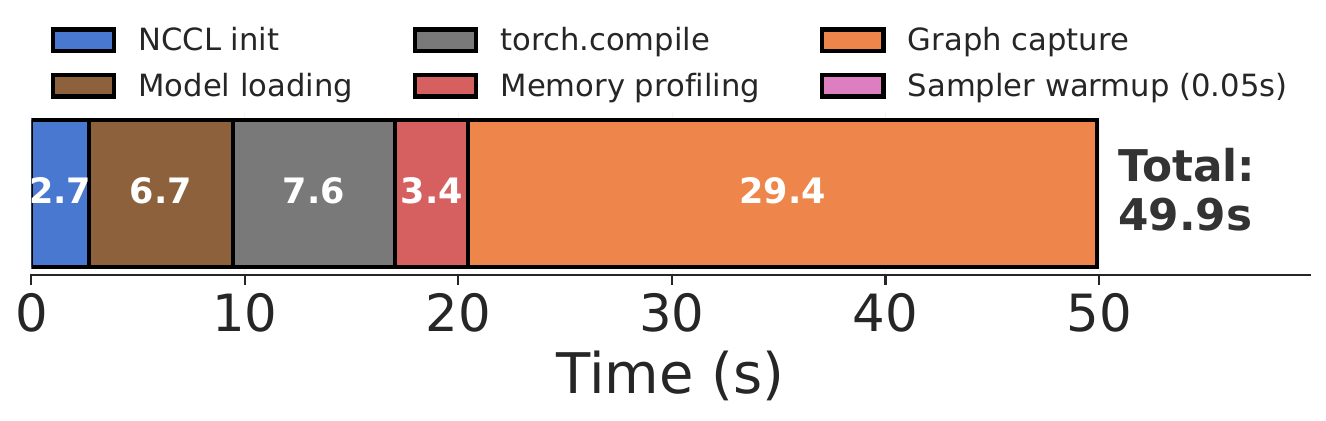}
\caption{Breakdown of vLLM worker initialization of serving Qwen3-14B on 2xH200s.}
\label{fig:init-breakdown}
\end{figure}

LLM service providers such as OpenAI~\cite{openai-models-docs}, as well as serverless LLM platforms such as Amazon Bedrock~\cite{amazon-bedrock}, typically rely on autoscaling to adjust the number of serving instances for each model in response to changing online demand. When demand spikes, the system launches new serving instances, whose cold-start latency directly increases the time-to-first-token (TTFT) of affected requests. As a result, cold-start latency often becomes the critical-path bottleneck of autoscaling in LLM serving.

The cold-start process of an LLM serving instance consists of two parts. The first is \textit{environment initialization}, which is largely model-agnostic and includes spawning Python worker processes, importing PyTorch/vLLM and their dependent libraries, and setting up generic host-side runtime state. Prior work shows that this cost is modest (typically under 10s) and can largely be removed from the critical path using pre-warmed workers that have already completed process creation and library import, but do not yet occupy model-specific GPU resources~\cite{zeng2025medusa,hu2025deepserve}. The second is \textit{worker initialization}, which constructs model- and deployment-specific serving state on the GPU, including communicator initialization (e.g., NCCL), model weight loading, compilation cache loading (e.g., \textit{torch.compile}), KV-cache profiling, and CUDA graph capture for efficient execution.

In \autoref{fig:init-breakdown}, we breakdown the worker initialization phase of vLLM. Most of the components have already been substantially optimized by prior work. Weight loading can be accelerated via RDMA-based transfer~\cite{fu2024serverlessllm,zhang2025blitzscale,zhu2025tangram,perplexity-rdma-weight-transfer,lmsys-rfork}, which enables a 1T parameter model to be loaded in under 2s~\cite{perplexity-rdma-weight-transfer}. KV cache profiling can be performed offline to eliminate its runtime overhead~\cite{zeng2025medusa}.

After these optimizations, CUDA graph capture remains the dominant bottleneck. As shown in \autoref{fig:init-breakdown}, vLLM already loads warmed \texttt{torch.compile} JIT caches rather than recompiling from scratch, yet it still takes 8s. In contrast, \sys bypasses even this overhead by directly materializing CUDA graphs together with the compiled kernel binaries embedded in the execution context, eliminating the need for \texttt{torch.compile}'s Python bytecode transformation.

Besides cold starts, recent work has explored dynamic parallelism reconfiguration~\cite{liu2025expert,gu2025elastic,mooncake2026elasticep,chen2025gyges} to further improve resource efficiency in LLM serving in response to workload dynamics. These approaches also suffer from the high cost of recapturing CUDA graphs after each parallelism switch.

\subsection{CUDA Graphs are Performance-Critical}
\begin{figure}[t]
\centering
\includegraphics[width=0.99\linewidth]{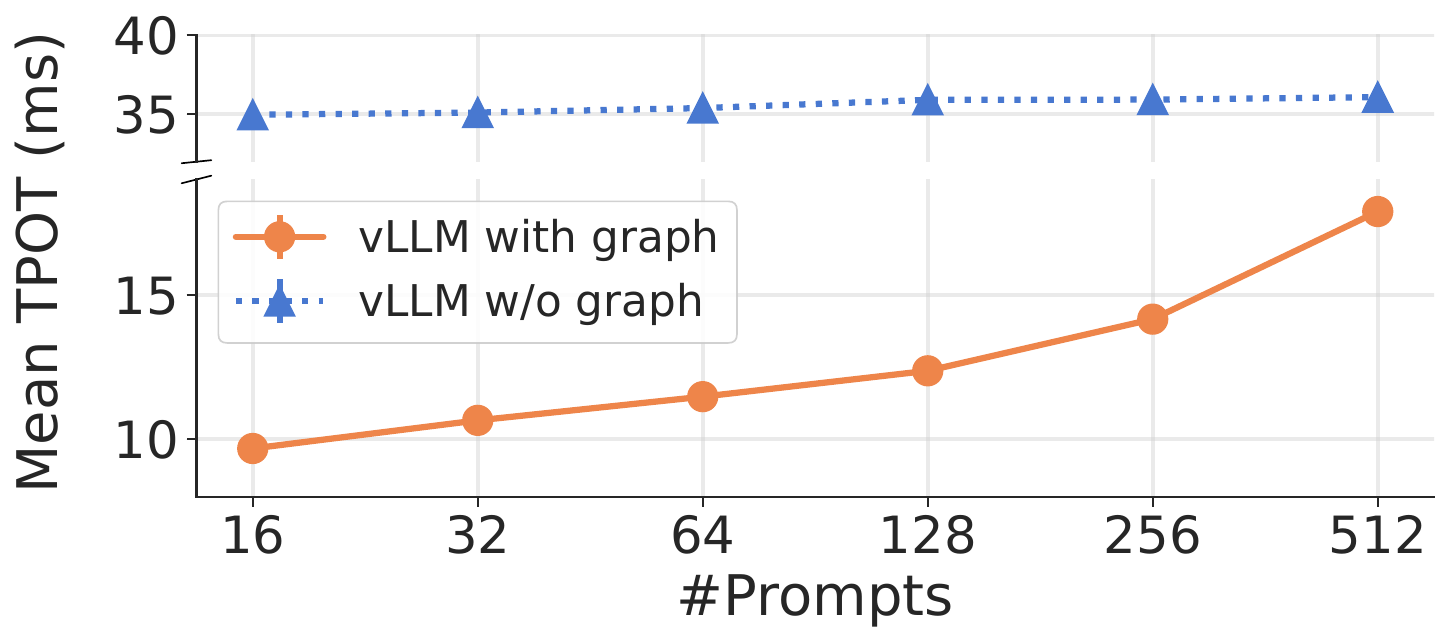}
\caption{Time per output token (TPOT) under different batch sizes for Qwen3-30B-A3B on 2xH200s with EP2, using vLLM with and without CUDA graphs.}
\label{fig:eager-vs-graph}
\end{figure}

CUDA graphs (and HIP graphs on AMD GPUs) provide a graph-based work submission model that represents a GPU workload as a directed graph, where nodes correspond to GPU operations and edges encode their execution dependencies. This abstraction allows a sequence of operations to be defined once and then launched repeatedly as a single execution unit. CUDA graphs have become a de facto mechanism for accelerating LLM inference because LLM decoding consists of many short-lived GPU kernels, especially as GPU architectures continue to improve. If these kernels are launched one by one from the CPU, host-side launch overhead can become a significant fraction of end-to-end latency. CUDA graphs mitigate this overhead by launching an entire graph of kernels at once, rather than issuing each kernel individually. As shown in \autoref{fig:eager-vs-graph}, disabling CUDA graphs noticeably degrades decoding performance.

CUDA graphs, however, come with a non-trivial initialization overhead. In modern inference engines, they are usually built through \textit{stream capture}, in which a regular model forward pass is executed while the CUDA stream is placed in capture mode. During this procedure, the CUDA runtime records all kernel launches and other CUDA operations, such as \texttt{memcpy}, along with the dependencies between them, into a graph. In practice, inference engines also perform several warmup forward passes before capture so that one-time initialization steps, including cuBLAS initialization and Triton autotuning, are completed in advance~\cite{vllm-cuda-graphs}.

\subsection{Existing Approaches and Their Pitfalls}
\begin{figure}[t]
\centering
\includegraphics[width=\linewidth]{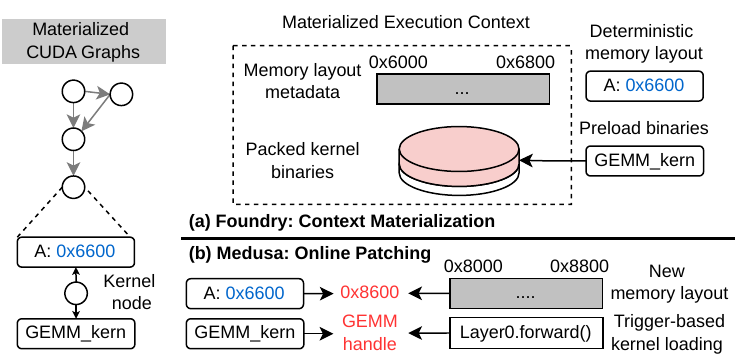}
\caption{Medusa materializes only the graph topology, while \sys also materializes the execution context.}
\label{fig:bg_medusa_compare}
\end{figure}

\begin{figure*}[t]
\centering
\includegraphics[width=\linewidth]{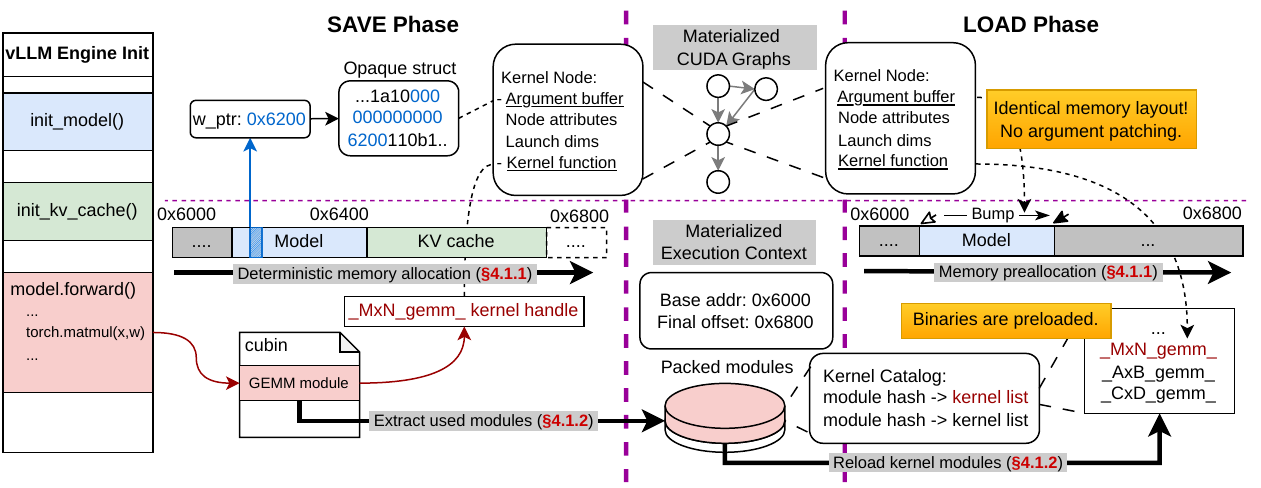}
\caption{Overview of \sys's workflow.}
\label{fig:overview}
\end{figure*}

Recently, Medusa~\cite{zeng2025medusa} proposed mitigating CUDA graph capture overhead by materializing graph topology during an offline capture phase and reconstructing graphs using CUDA’s explicit graph construction APIs~\cite{nvidia-cuda-graph-api}. As shown in \autoref{fig:graph-build-time}, graph construction via these APIs is 2--3$\times$ faster than stream capture. However, a CUDA graph is defined not only by its topology, but also by the execution context in which it was captured, whereas Medusa materializes only the former, as shown in \autoref{fig:bg_medusa_compare}. In particular, each kernel node contains a handle to the corresponding kernel function, while its argument buffer may embed pointers to GPU-resident objects such as model weights, the KV cache, and activations. To restore a graph during online serving, Medusa applies hand-crafted, kernel-specific patching rules to rewrite these pointer arguments so that they refer to memory allocated in the current run. To recover kernel handles, Medusa still executes and captures the first layer of the model for each batch size, thereby triggering the loading of the relevant kernel modules. This reliance on hand-crafted patching rules makes Medusa difficult to generalize to new models and hardware platforms, where new kernels are introduced frequently. Moreover, its design becomes increasingly fragile as models and kernel libraries evolve. For example, newer versions of cuBLAS pack pointer arguments into opaque structures, making them difficult to identify and patch (see \autoref{app:kernel_node_example}). In addition, many recent models use a dense architecture in the initial layers and MoE architecture thereafter, invalidating Medusa’s assumption that all layers are structurally identical and that the first layer suffices to trigger the loading of all relevant kernels.

Meanwhile, process-level checkpoint/restoring (C/R) is recently supported by CUDA driver~\cite{nvidia2025cudacheckpoint,stoyanov2025criugpu}, which can be used in conjunction with CRIU~\cite{criu_github} to checkpoint and restore the worker process along with its CUDA states. However, this approach tightly couples CPU and GPU state at the process level, making it ill-suited to dynamic parallelism switching, where request and KV-cache state must be preserved while the parallelism configuration changes. In addition, the CUDA driver’s C/R functionality suffers from slow restoration and does not yet support the IPC memory required for multi-GPU inference. Beyond driver-integrated C/R, many interception-based C/R techniques~\cite{zeng2026gpu,wei2025phoenixos} have also been proposed. These systems intercept CUDA driver API calls, record resource handles, and restore them via API replay. However, none of them supports CUDA graphs.

Furthermore, as a single serving instance has scaled to hundreds of GPUs with the popularity of MoE models~\cite{deepseek_open_infra_2025,zhao2025insights}, it is infeasible to frequently allocate a large number of GPUs solely for offline processing. Even if Medusa and C/R-based approaches apply to multi-GPU scenarios, they still suffer from high processing cost, as they need to prepare a materialized graph or checkpoint for all worker processes, even when those workers share the same computation logic.

%% file: overview.tex
\section{Overview}
\label{sec:overview}

We propose \sys, a system that persists captured CUDA graphs together with the \emph{execution context} they depend on during a one-time offline capture run, and restores them in a fresh serving instance at negligible cost.

A captured CUDA graph is not self-contained: its kernel nodes may embed device addresses that are valid only in the capturing run, and may reference kernels that are lazily loaded during warmup. These context-dependent references make captured graphs non-portable to fresh processes.

\sys addresses this portability problem by materializing the execution context required for graph replay. At a high level, it re-establishes the two conditions that make captured graphs replayable in a fresh process: the memory layout expected by captured kernel arguments, and the kernel functions referenced by graph nodes. Concretely, \sys enforces a deterministic memory layout by interposing on the CUDA driver's virtual memory management (VMM) APIs (\S\ref{sec:design-mem}), and restores the required kernel functions by automatically extracting and reloading the kernel binaries used by the captured graphs (\S\ref{sec:design-binary}). As a result, \sys remains library-agnostic and does not require inspecting or patching kernel-specific argument layouts.

As shown in \autoref{fig:overview}, \sys operates in two phases: \save and \load. During \save, \sys runs the engine’s normal warmup and graph capture once, while intercepting CUDA driver calls to record the graph-dependent state required for replay. The output is a portable \workspace that packages serialized graph metadata together with the execution-context information needed for reconstruction, including memory-layout metadata and kernel binaries. During \load, each serving process consumes this \workspace to restore the execution context and reconstruct executable graphs, without re-executing warmup or graph capture.

\sys reduces \load cost by exploiting a key structural regularity of CUDA graphs: graphs captured for different batch sizes often share the same \emph{topology} and differ only in \emph{per-node parameters}. Concretely, the topology includes node types, dependency edges, and other attributes like cluster dimensions, while the per-node parameters include kernel arguments and launch dimensions, i.e., \texttt{gridDim} and \texttt{blockDim}. Since the CUDA driver supports updating per-node parameters on a constructed graph without re-instantiation, \sys builds only one template graph for each unique topology instead of reconstructing every captured graph from scratch. Inference engines may capture hundreds of graphs, but these typically collapse to only a small number of unique topologies (e.g., 22 for Qwen3-14B on a H200). \sys then specializes the template on demand at serving time by applying the target per-node parameters. (\S\ref{sec:design-grouping}).

\sys further extends the \emph{templating} insight to distributed LLM serving. In SPMD-style parallelism, such as DP, TP, and EP, all ranks follow the same computation flow and thus share the same graph topology, differing only in specific model shards and rank-dependent communication state. \sys exploits this invariance by performing \save on a single GPU with dummy communication, and then reconstructing rank-specific graphs during \load by patching in the actual communication handles and rank identifiers. As a result, a single offline capture can be reused across all ranks in a distributed deployment, eliminating redundant warmup runs and reducing \workspace storage proportionally, e.g., by 64$\times$ for a 64-GPU cluster. (\S\ref{sec:design-forging}).

In summary, \sys performs \save once on a single GPU to materialize graph templates and the execution context required for graph replay, and produce a portable \workspace. During serving, each rank runs \load to restore executable graphs from the templates. This design allows offline-captured templates to be reused across batch sizes and distributed ranks, reducing both cold-start latency and offline processing cost.

%% file: design.tex
\section{Design}
\label{sec:design}

\subsection{Execution Context Materialization}
\label{sec:design-restore}

A captured CUDA graph is not self-contained: its kernel arguments may contain device addresses that are valid only in the capturing run, and its kernel nodes may reference functions loaded during the pre-capture warmup runs for each batch size. In a normal engine startup, both dependencies are satisfied implicitly as a side effect of initialization: model and KV-cache initialization allocate the required device state, and warmup execution triggers the loading of the required kernel modules. \sys makes these dependencies explicit by materializing the execution context required for graph replay, so that \load can directly reconstruct executable graphs in a fresh process without re-executing warmup.

\begin{figure}[t]
\centering
\includegraphics[width=\linewidth]{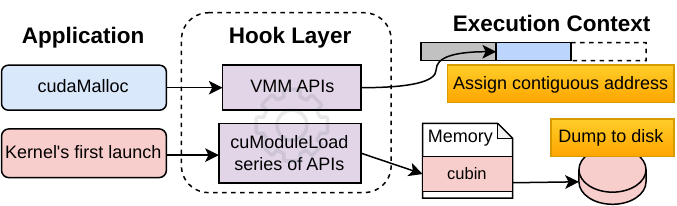}
\caption{\sys works as an interposition layer by implementing a CUDA driver hook.}
\label{fig:hook}
\end{figure}

\subsubsection{Deterministic Memory Layout}
\label{sec:design-mem}

Nodes in a captured CUDA graph embed memory pointers in their function arguments, 
and some CUDA libraries (e.g. cuBLAS) further hide pointers inside opaque
structs passed as flat byte arrays (\autoref{app:kernel_node_example}). On a second run, 
CUDA driver may return different memory addresses even for the same program, and thus
all of these memory pointers are stale~\cite{zeng2025medusa}.  
As discussed in \autoref{sec:background}, patching embedded pointers to the new addresses
requires per-library knowledge of private parameter layouts and does not
generalize. \sys instead employs deterministic memory allocation mechanism to ensure 
data required by the graphs is at the same location on each run.

\sys resolves the problem by interposing the CUDA driver's virtual-memory management (VMM) interface. 
All CUDA allocation requests are redirected into a reserved virtual-address region
starting at a fixed base address chosen big enough to avoid conflicts. 

For LLM inference engines, these pointers in the graph usually refer to memory for model weights, the KV cache pool, and the input/output cache.
Those are large, long-lived objects that are allocated once and never freed throughout the serving lifetime~\cite{kwon2023pagedattention}. 
Besides, framework-level caching allocators retain reserved memory rather than returning it directly to the driver~\cite{pytorch-cuda-semantics}. 
\sys therefore exploits this monotonic nature, by placing the address of each allocation contiguously after the previous one during \save. 
Given the same base and the same allocation sequence, this mechanism deterministically
produces an identical and continuous address layout. No kernel argument buffer needs to be
inspected or rewritten. Because the actual device memory free call during initialization is rare, the fragmentation waste is negligible. 

During \load, because \sys skips the graph capture to save time, the transient intermediate buffer allocations during the graph capture window will not occur,
while their addresses are also embedded in the captured graph. Therefore, the allocation sequences between \save and \load are not perfectly identical. 
Because \sys tracks memory event, it handles this problem by recording all capture-window allocations during \save and
replaying them during \load, ensuring that the full address space expected by
the graph is present (\S\ref{sec:impl}). This is also why simply disabling
address-space layout randomization (ASLR)~\cite{linux-proc-sys-kernel-aslr} would not suffice, as the allocation
sequences themselves differ between \save and \load.

The monotonic allocation strategy also enables a performance optimization via \emph{preallocation},
which is critical to efficient graph reconstruction. Each virtual-memory allocation
requires mapping an address range and setting access permissions, and repeated
calls to allocate fine-grained buffers introduce significant
overhead~\cite{prabhu2025vattention}. Because initialization under \sys will produce a continuous memory layout, 
the final offset recorded during \save fully describes its
extent. \load can therefore map the entire range up to that offset in a single
allocation before any individual allocation occurs. All subsequent allocation
requests, including model weights and KV cache initialization, and the capture-window buffer replay, 
simply verify that the requested range falls within the
preallocated region and advance the pointer offset, reducing each allocation to
a nanosecond pointer bump. This also ensures that every device address
referenced by graph nodes is live before graph construction begins, which also allows the template-based reconstruction
\S\ref{sec:design-grouping} runs asynchronously from other initialization tasks.

\subsubsection{Binary Extraction and Reload}
\label{sec:design-binary}

Each kernel node in a CUDA graph holds a reference to a kernel function that performs the
actual GPU computation. Kernel functions are organized into a \texttt{CUmodule} (or \texttt{CUlibrary} introduced in CUDA 12\footnote{NVIDIA CUDA Driver API, ``Library Management,'' \url{https://docs.nvidia.com/cuda/cuda-driver-api/group__CUDA__LIBRARY.html}.}), including proprietary libraries such as cuBLAS. These modules enter the CUDA context via \texttt{cuModuleLoad} (or \texttt{cuLibraryLoad}) series of APIs. 
By default, these modules are lazily loaded when a kernel is first launched or explicitly
requested. A process that skips warmup never triggers these loads, so directly replaying 
the graph fails due to the absence of the required kernel functions.

\sys resolves this issue by intercepting those load APIs during \save to track which modules/libraries are loaded and restore them at the beginning of \load.
During \save, \sys extracts the code objects of used modules/libraries from the in-memory shared libraries (e.g., \texttt{libcublas.so}) 
and records: (1)~the binary payload of a module/library, (2)~the load API and options
used to load it (so \load can replay the same driver call), and (3)~a catalog
that maps each binary to its contained kernel entry points, keyed by the
binary's content hash and each kernel's mangled function name. During \load, these binaries are
restored into the driver using the recorded load paths. When graph
reconstruction encounters a kernel node, it resolves the node's function
reference by looking up the same (hash, name) key in the catalog, obtaining a
valid kernel function handle without depending on a warmup execution to lazily
reload the kernel. Compilation (e.g., \texttt{torch.compile}) and autotuning work paid during cold start are also avoided.

Some binaries require additional preparation before they can be loaded. Certain
CUDA libraries produce device code as multiple relocatable segments that
must be linked into a single binary before they can be loaded as a module.
Repeating this linking at \load time would add unnecessary latency. \sys instead
pre-links the segments into a ready-to-load cubin during \save. Additionally, some modules require
post-load initialization of device-side runtime state (e.g., DeepEP's
communication kernels depend on NVSHMEM, which must initialize device-side state
pointers inside each loaded module). \sys detects such requirements during \save and records a flag in the
\workspace to avoid probing every loaded module during \load.     

\subsection{Efficient Graph Reconstruction}
\label{sec:design-low-cost}

With deterministic memory layout and preloaded binaries, \sys can reconstruct replayable
graphs from serialized metadata. To scale graph reconstruction in realistic distributed
LLM serving, \sys must address two bottlenecks. Within a
single rank, inference engines typically capture one CUDA graph per
supported batch size to minimize latency, and rebuilding hundreds of graphs through the CUDA driver is expensive.
Across ranks, na\"ively producing the \workspace for an entire cluster requires
running \save on every rank, consuming the full hardware allocation and storing
redundant copies of graphs and binaries. \sys addresses both through \emph{templating}.
Because captured graphs share identical topology and differ only in
per-node parameters, a small number of templates can represent the entire graph set.

\begin{figure}[t]
\centering
\includegraphics[width=\linewidth]{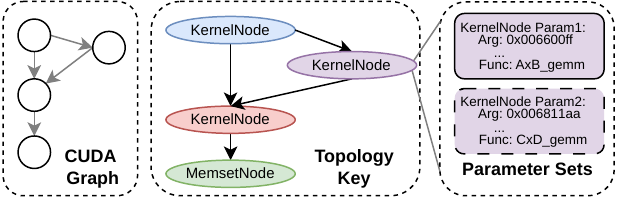}
\caption{CUDA graph topology and node parameter sets.}
\label{fig:graph-topo}
\end{figure}

\subsubsection{Topology-based Intra-Rank Graph Grouping}
\label{sec:design-grouping}

Reconstructing a CUDA graph incurs many CUDA driver calls, including node addition,
attribute setting, and graph instantiation, and the total number of such calls scales with model size and graph complexity. Although NVIDIA claims that graph objects 
are not internally synchronized\footnote{Graph object thread safety,  \url{https://docs.nvidia.com/cuda/archive/12.8.1/cuda-driver-api/graphs-thread-safety.html}.}, 
we observe significant contention on driver calls even when building independent graphs on separate
threads. The per-call latency increases with thread count and total wall-clock time
barely improves (\autoref{app:parallel_graph_build}). Simply parallelizing graph
construction is therefore neither safe nor effective.

After analyzing the CUDA graphs for multiple LLM architectures, we realized that 
graphs captured for nearby batch sizes share the same topology. 
They share identical node types in the same order with the same dependency structure, and differ only in per-node parameters (\autoref{fig:graph-topo}). 
For example, among CUDA graphs for Qwen3-14B~\cite{qwen3-14b-model} captured for batch size 1--512 on an H100, there are only 22 unique topologies.
On the other hand, we found out CUDA driver provides APIs to update per-node parameters of an instantiated graph executable~\cite{nvidia-cuda-graph-api}, 
and it runs up to 20x faster than assemble and instantiate a new graph (\S\ref{eval:on-demand}).
Therefore, rather than building every graph during \load, \sys builds
only one \emph{template graph} per distinct topology and prepares parameter sets
of all other graphs in the same group that can be applied to the template on demand.

\sys identifies unique topologies during \save to reduce the reconstruction workload of \load. After capturing all
graphs, \sys computes a topology key for each graph that encodes its structural and topological
properties in a compact form. Graphs with identical keys are grouped together, and the resulting grouping information is recorded in a manifest for later reconstruction.

During \load, \sys only constructs template graphs via CUDA driver APIs.
Non-template graphs require only node parameter preparation, which
involves no driver calls and its speed scales linearly with CPU threads. Because 
\sys directly determines the grouping via the manifest, template
construction and on-demand parameter preparation run \emph{concurrently}: multiple worker
threads prepare non-template graphs in parallel while the main thread builds
templates sequentially to avoid driver contention. The details of topology grouping and on-demand update mechanism are described in
\S\ref{sec:impl}. Furthermore, because the memory are preallocated at the beginning of \load (\S\ref{sec:design-mem}), 
addresses embedded in the graph are already valid, and thus the whole graph reconstruction 
can run asynchronously with the foreground initialization (e.g. model weights, kv cache and capture window buffers).

At replay time, if the target graph's batch size matches that of template graph, 
\sys replays it directly. Otherwise, \sys applies the
node parameters of the target graph to update the template executable before replay.
This is a one-time effort, as repeated calls to the same graph skip the update entirely.
We demonstrate that such on-demand specialization is lightweight in \S\ref{eval:on-demand}.

\subsubsection{Inter-Rank Graph Sharing}
\label{sec:design-forging}

In SPMD-style parallel deployment, such as data parallelism (DP), tensor parallelism (TP), and expert parallelism
(EP), the computation is invariant across ranks, while the communication kernels differ in
rank-specific arguments (e.g., the rank identifier passed to collective kernels). 
The shared execution flow across participating GPUs ensures that they load identical kernel binaries, while \sys's deterministic allocation mechanism enforces a consistent memory layout. 
Together, these properties allow \sys to capture a graph on a single GPU, then share it across all participating GPUs. 
This reduces both the hardware cost of \save and the \workspace footprint from cluster-scale to single-device. In practice, operators often
maintain \workspace{}s for multiple parallelism configurations to meet varying
latency and throughput targets, incurring even higher cost.

Because a single GPU cannot replay the full collective communication operation without peers, we cannot run the original communication kernels during capture.
However, our insight is that the communication kernels rely on common building blocks, such as NCCL and NVSHMEM. 
Therefore, we can build a stub over these common building blocks and perform a dummy communication during capture.
During \load, \sys substitutes the handle with the real kernel function and updates the rank-specific arguments.

Inter-rank graph sharing applies whenever all ranks follow the same execution flow and thus produce equivalent graph structure during capture.
Pipeline parallelism (PP) is not included, since different ranks execute different model stages and therefore require stage-specific graph capture.
Moreover, PP introduces inter-stage bubbles that increase per-request latency during
decoding~\cite{agrawal2024sarathiserve}, making it less suitable for latency-sensitive serving workloads.
\sys can also support hybrid strategies (e.g., EP combined with TP) as long as the resulting combination preserves topology invariance across ranks and does not introduce rank-specific execution divergence.

%% file: implementation.tex
\section{Implementation}
\label{sec:impl}

\sys is implemented as two components: (i)~a \emph{driver-hook library}, a
C/C++ interposition layer injected via \texttt{LD\_PRELOAD} that enforces
deterministic memory allocation and captures kernel binaries, and (ii)~a
\emph{graph extension} to PyTorch, a C++ library with Python bindings that serializes,
reconstructs, and replays CUDA graphs with template-based sharing. A thin
Python integration layer wires these components into the vLLM~\cite{kwon2023pagedattention} to demonstrate LLM serving.
The driver-hook library and graph extension together comprise approximately
10K lines of C++ and 2K lines of Python.

\subsection{Memory Allocation Redirection}
\label{sec:impl-mem}

The deterministic allocation mechanism (\S\ref{sec:design-mem}) is
implemented by redirecting CUDA memory allocation to a virtual memory management (VMM) backend. 
VMM can map a physical memory buffer onto a specified virtual address, as long as it is not occupied.
\sys reserves a large virtual address space at the beginning. For each following 
allocation, \sys leverages VMM to assign the virtual address right after the last one, and results in a deterministic and contiguous memory region.
Because \sys tracks all incoming allocation calls, it can record and replay a selected range of memory events as needed.

\subsection{CUDA Module Load Interception}
\label{sec:impl-binary}

The extraction of kernel binaries (\S\ref{sec:design-binary}) is implemented
by intercepting module load driver APIs. \texttt{CUmodule}s, which contain kernel handles used by CUDA graphs,
are loaded via \texttt{cuModuleLoad} and its series of APIs. During \save, 
\sys works as a wrapper to track the data pointer passed into the API
and extracts its raw binary payload from the memory and computes a hash.
\sys also enumerates through all the kernel handle entrypoints of loaded modules and records a catalog along with the binary payload.
Therefore, during \load, it can directly resolve the kernel handle from the kernel function name, 
instead of searching through all the loaded modules. Some modules such as DeepEP use NVSHMEM that requires device-side state initialization after loading.
\sys detects that during \save and sets a flag, and calls \texttt{nvshmemx\_cumodule\_init} \footnote{NVSHMEM APIs, \url{https://docs.nvidia.com/nvshmem/api/gen/api/setup.html\#nvshmemx-cumodule-init}} on it during \load.

\subsection{Graph Grouping and Serialization}
\label{sec:impl-graph}

\sys groups hundreds of graphs into a few templates to save time on reconstruction. The on-demand graph update is implemented
via \texttt{cuGraphExecUpdate}, which can update node parameters, 
such as kernel function, argument buffer, and launch dimensions, without re-instantiating the graph executable object.
It requires the new graph to have the same topology as the template, including their node types, order, and dependencies between nodes.
According to our test, kernel node attributes (e.g., cluster dimensions) set via \texttt{cuGraphKernelNodeSetAttribute}, also remain stale, 
and thus it is treated as a part of graph topology. \sys computes a key based on topology and organizes graphs with the same key into a group.
Across different model architectures and sizes, graph grouping effectively reduces the number of graphs to build during \load (\S\ref{sec:eval-template-eff}).

We initially saved each cuda graph structure into a JSON file for better readability. However, as model size increases, the parsing incurs significant delay.
As a result, \sys also serializes a binary version, which effectively reduces the parsing time of 512 graphs from a few seconds to less than 100 milliseconds.

\subsection{LLM Serving Engine Integration}
\label{sec:impl-vllm}

Because CUDA graphs are mainly designed for LLM decoding to eliminate the bubbles caused by host-side delay, 
our optimization target is the decoder of prefill-decode disaggregated serving~\cite{distserve}. 
To demonstrate multi-GPU capability, \sys is evaluated on expert-parallel (EP) serving, which is the 
state-of-the-art LLM deployment strategy for frontier mixture-of-expert (MoE) models~\cite{deepseek_open_infra_2025}.
Besides the specialized communication kernels in DeepEP, \sys seamlessly supports FP8 expert computation (DeepGEMM), thanks to its library-agnostic design.

We integrate \sys into vLLM to accelerate the engine initialization, by implementing a thin layer under its compilation model wrapper to apply our graph extension, 
and preload our hook layer when each worker process is created. The kv cache size of vLLM can change because it is set via the utilization percentage of currently available memory. 
We specify the kv cache size before \save or \load so that the memory layout is consistent across different runs. 
Because memory is preallocated at the beginning of \load (\S\ref{sec:design-mem}), the graph construction process works asynchronously from the foreground tasks such as kv cache initialization.

%% file: evaluation.tex
\section{Evaluation}
\label{sec:eval}

\begin{figure*}[t]
\centering
\includegraphics[width=\linewidth]{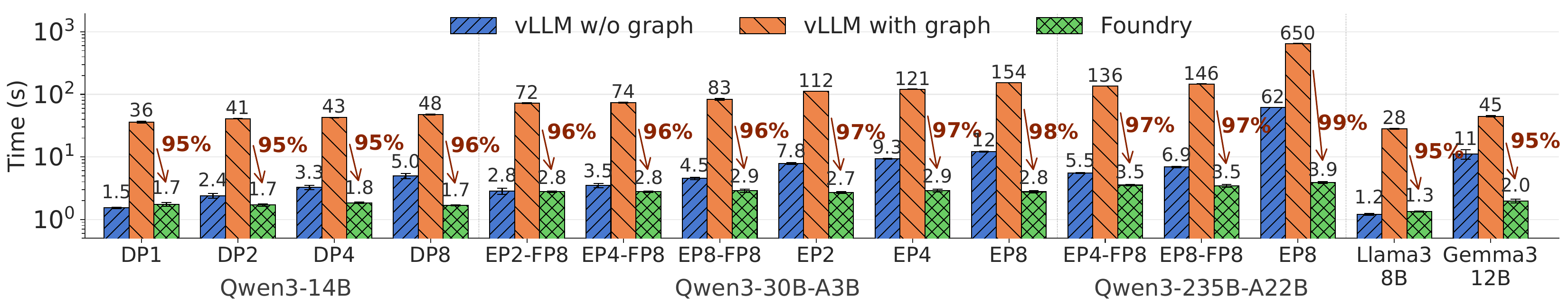}
\caption{\textbf{[Cold-Start Latency Reduction]} Serving engine initialization latency across models and parallelism configurations on H200. Percentages indicate \sys's reduction relative to vLLM with CUDA graphs.}
\label{fig:init-all-model}
\end{figure*}

\begin{figure}[t]
\centering
\includegraphics[width=\linewidth]{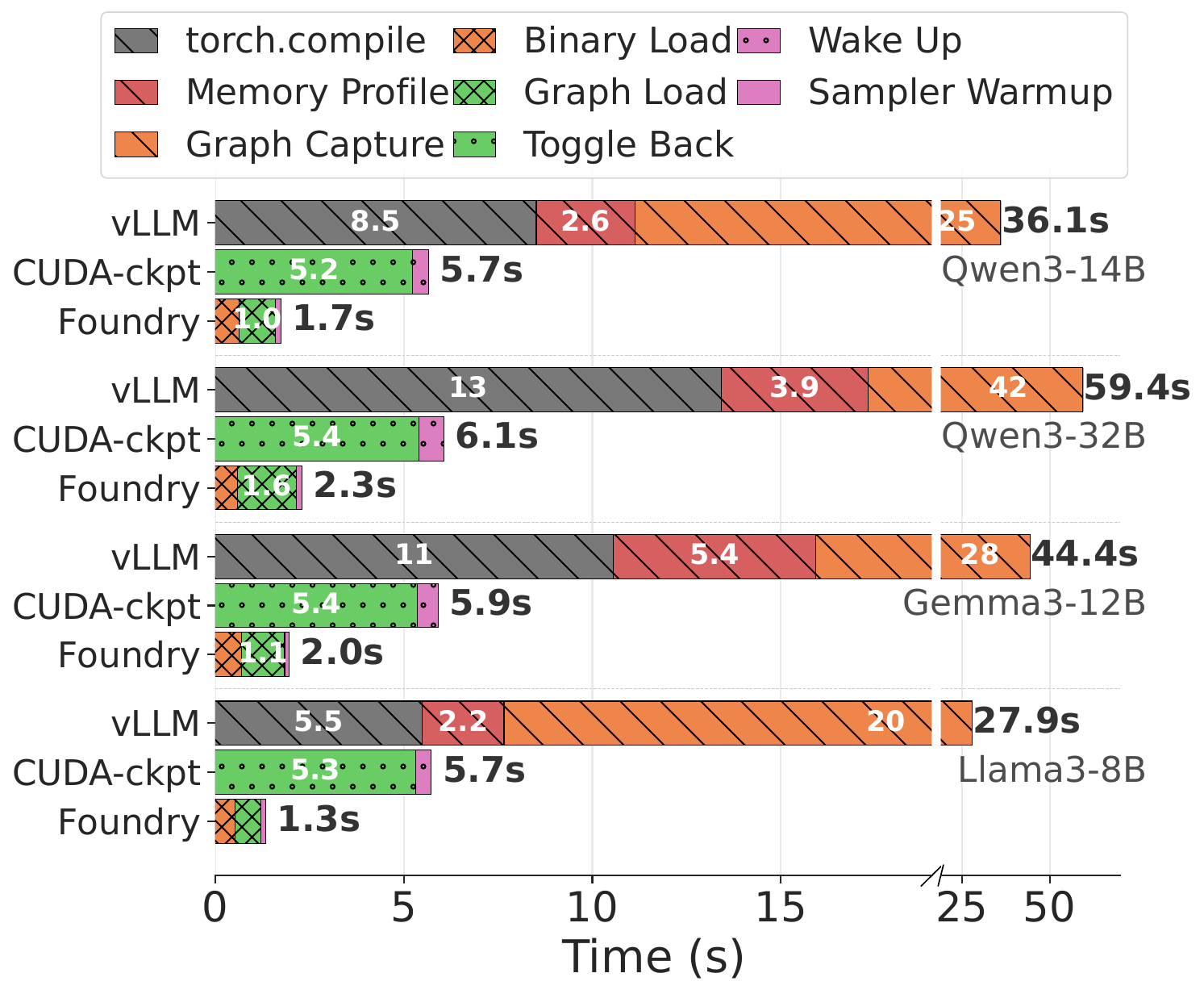}
\caption{\textbf{[Cold-Start Latency Reduction]} End-to-end engine initialization breakdown comparing vLLM, CUDA-checkpoint, and \sys across four model configurations. Note: \texttt{torch.compile} only does bytecode transformation and cache loading.}
\label{fig:cuda-ckpt}
\end{figure}

\begin{figure*}[t]
\centering
\includegraphics[width=\linewidth]{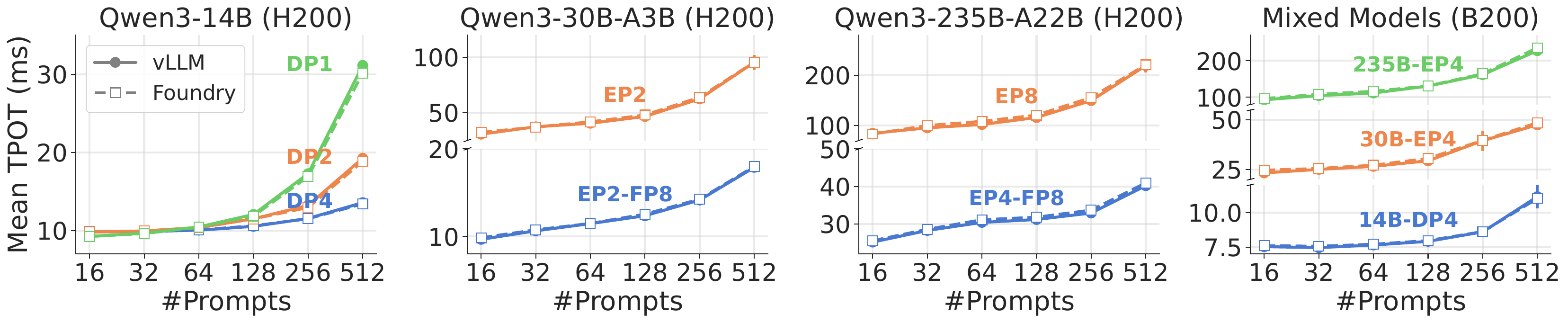}
\caption{\textbf{[Serving Throughput Preservation]} Mean time per output token (TPOT) for vLLM and \sys across batch sizes 16--512.}
\label{fig:eval-tpot}
\end{figure*}

\begin{figure}[t]
\centering
\includegraphics[width=\linewidth]{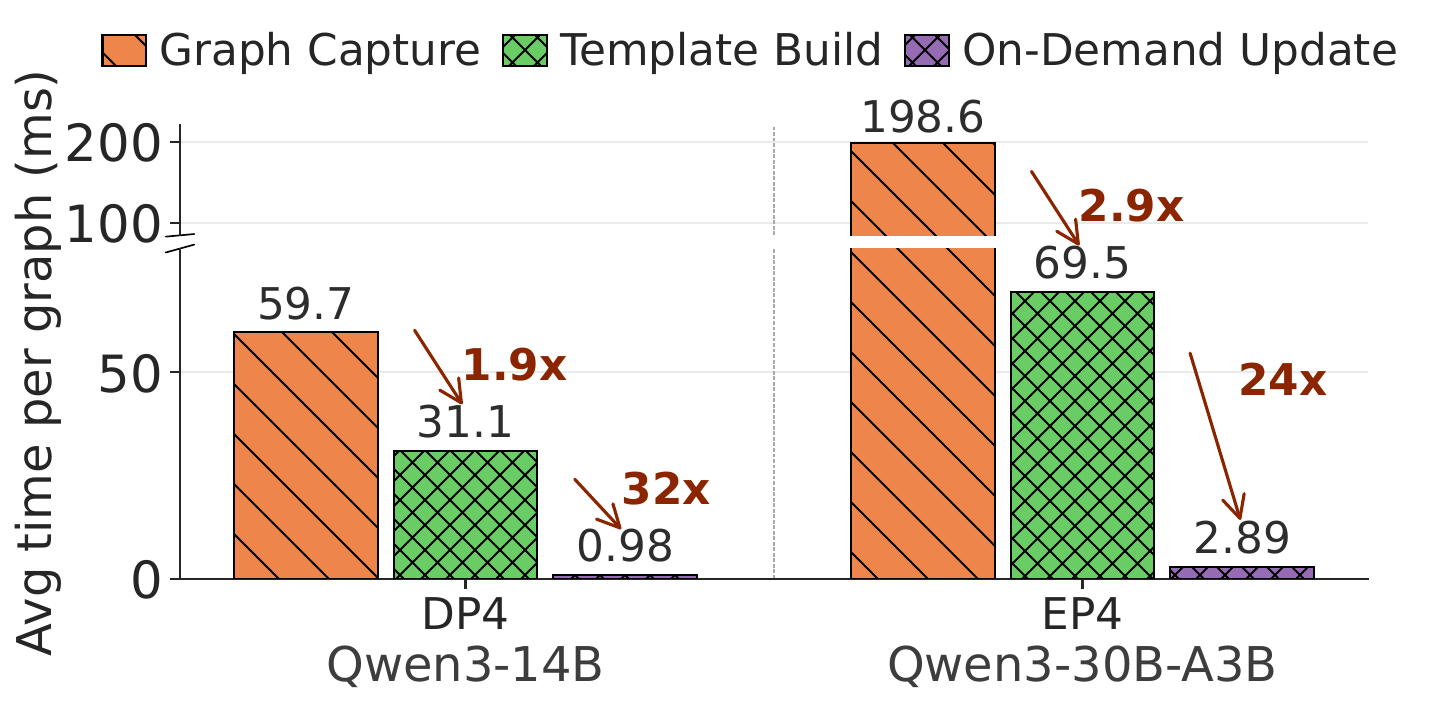}
\caption{\textbf{[Effectiveness of Templating]} Average per-graph cost of stream capture, template construction via driver APIs, and on-demand parameter update. All experiments capture 512 graphs.}
\label{fig:graph-build-time}
\end{figure}

\begin{figure}[t]
\centering
\includegraphics[width=\linewidth]{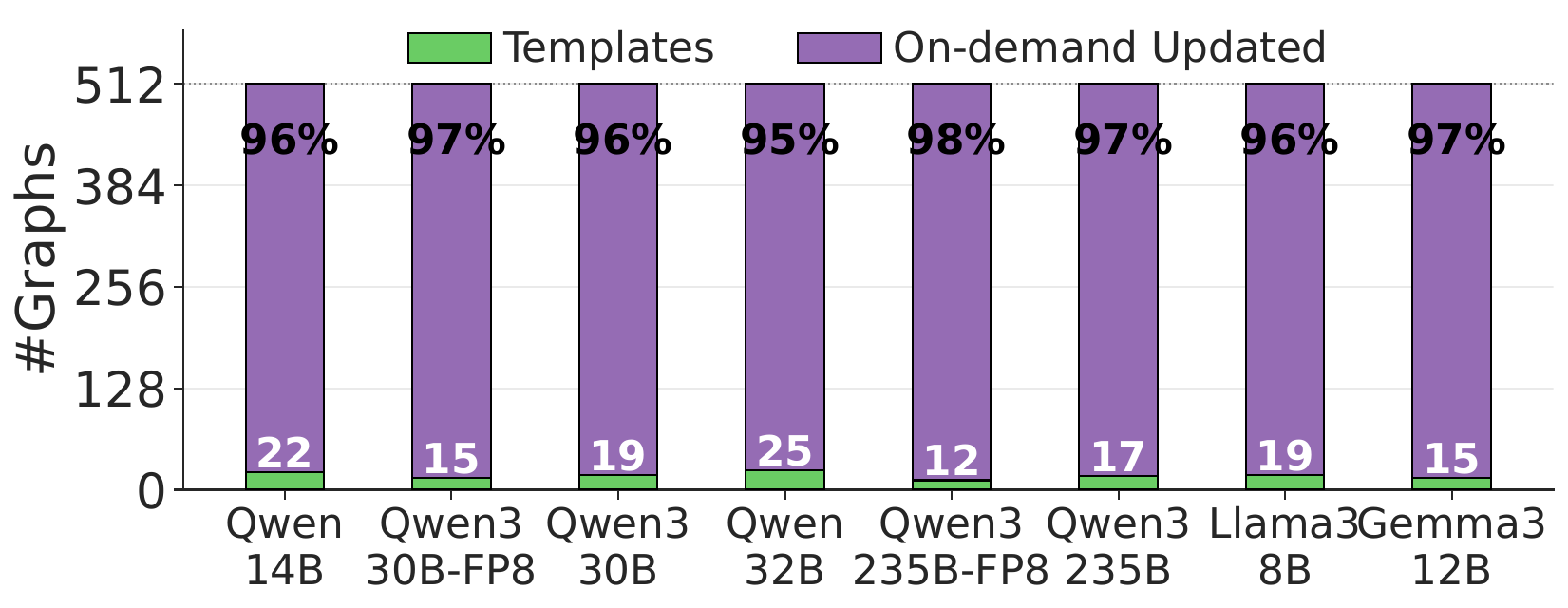}
\caption{\textbf{[Effectiveness of Templating]} Number of unique templates versus on-demand updated graphs out of 512 total. Percentages indicate the fraction of graphs served via on-demand update.}
\label{fig:template-percentage}
\end{figure}

We evaluate \sys along four dimensions:
(1)~reduction in cold-start latency compared to vanilla vLLM and CUDA-checkpoint;
(2)~preservation of serving throughput;
(3)~effectiveness of template-based graph reconstruction; and
(4)~storage cost saving.

\subsection{Experimental Setup}
\label{sec:eval-setup}

\paragraph{Testbed Configurations.}
All experiments run on NVIDIA DGX nodes. The primary platform uses 8$\times$H200 GPUs with Intel Xeon Platinum 8480C CPUs and 2\,TB host memory. We also evaluate on a second node with 8$\times$B200 GPUs and Intel Xeon Platinum 8570 CPUs to demonstrate \sys performance on the latest architecture. All GPUs in a single instance are fully connected with NVLink. 

Both nodes run NVIDIA driver 590.48 with CUDA 13.1. We use vLLM v0.11.2, PyTorch 2.9 and NVSHMEM 3.3.24 in our experiment environment. For all experiments, we load the warmed cache of \texttt{torch.compile} and \workspace of \sys from local NVMe storage.

\paragraph{Models and parallelism.}
We evaluate across three model families covering a wide range of architectures and sizes:
Qwen3-14B~\cite{yang2025qwen3}, Qwen3-32B, Llama3-8B~\cite{grattafiori2024llama} and Gemma3-12B~\cite{kamath2025gemma} are  dense models, while
Qwen3-30B-A3B and Qwen3-235B-A22B are mixture-of-experts (MoE) models.
Dense models are deployed with data parallelism (DP1--DP8); MoE models with expert parallelism (EP2--EP8), in both BF16 and FP8 precision. All configurations capture 512 CUDA graphs covering batch sizes 1--512 to simulate real world deployment. 

\paragraph{Baselines.}
We compare against two baselines: (1)~\emph{vLLM with CUDA graphs}, the default production implementation that performs full warmup and stream capture during initialization; and (2)~\emph{CUDA-checkpoint}~\cite{nvidia2025cudacheckpoint}, which snapshots the entire CUDA context via NVIDIA's process-level checkpoint/restore API, used in conjunction with CRIU~\cite{criu_github}.
We also report \emph{vLLM without CUDA graphs} (eager mode) as a reference for the minimum achievable startup time at the cost of degraded serving performance (\S\ref{sec:background}).

CUDA-checkpoint does not support the IPC memory required by communication kernels (e.g., DeepEP)~\cite{nvidia2025cudacheckpoint}, 
and its restore latency grows disproportionately for multi-GPU data-parallel engines, making it less efficient than launching multiple independent single-GPU instances.
We therefore compare with it only on single-GPU settings. To assess its full performance, we let vLLM releases model weights and the KV cache before checkpointing to avoid data loading bottlenecks.

We do not include Medusa~\cite{zeng2025medusa} because its per-kernel patching rules are implemented for a specific set of kernels and are based on an outdated CUDA driver version (550.54.14). Porting them to newer hardware and kernel libraries would require reimplementing the system. 
On Hopper and later GPUs, cuBLAS kernels use large argument buffers (\autoref{app:kernel_node_example}) with opaque layouts, making such patching impractical.

\paragraph{Metrics.}
We use cold start time to evaluate \sys{}'s ability to reduce initialization latency and use time per output token (TPOT) to measure the serving throughput, as we target the decoding phase of PD-disaggregated serving.   

\subsection{Cold Start Latency Reduction}
\label{sec:eval-cold-start}

\subsubsection{Comparison with vLLM}
\label{sec:eval-vllm}

\autoref{fig:init-all-model} compares engine initialization latency across 15 configurations spanning three model families on H200, including both dense and MoE models. For each configuration, we report the cold start latency of three modes: vLLM without CUDA graphs, vLLM with CUDA graphs, and \sys. Following Medusa~\cite{zeng2025medusa}, for all compared methods, we assume a pool of warm execution environments that eliminate the environment initialization overhead. We also omit weight loading time, as recent systems~\cite{lmsys-rfork,zhang2025blitzscale,yu2025lambda} have reduced it to 1\textasciitilde{}2s.

\paragraph{Dense models.}
For Qwen3-14B, vLLM with graphs takes 36--48\,s depending on the parallelism degree (DP1--DP8). Comparing with vLLM without graphs, graph capture clearly contributes the majority of the cost. \sys reduces this to 1.7--1.8\,s, a 95\% reduction that is consistent across all DP configurations. Notably, \sys's initialization time remains nearly constant as the number of ranks scales from 1 to 8. This is because both execution context restoration and CUDA graph reconstruction run fully independently, and there is no computation on GPUs. The reduction is generalizable to different model architectures, for example, on Llama3-8B and Gemma3-12B, the reduction is 95\% (28\,s$\to$1.3\,s) and 95\% (45\,s$\to$2.0\,s), respectively.

\paragraph{MoE models with default BF16.}
\sys exhibits significant latency reduction on large scale MoE models. Qwen3-30B-A3B (EP2--EP8) takes 112--154\,s to initialize on vLLM with graphs, reduced to 2.7--2.8\,s by \sys (97--98\%). The most dramatic case is Qwen3-235B-A22B EP8, where vLLM requires 650\,s (about 10 minutes) for initialization, of which graph capture alone accounts for the majority. \sys reduces this to 3.9\,s, a 99\% reduction. This extreme speedup arises because graph capture requires model forward on every supporting batch size, and a larger model tends to incur longer forward duration due to computation complexity. The latency of \sys also increases compared with small models as each graph contains more nodes, but it remains fast as it only builds a few templates and graph reconstruction is much faster than stream capture (\autoref{fig:graph-build-time}).

\paragraph{MoE models with FP8 quantization.}
By applying FP8 quantization, the vLLM initialization time of both Qwen3-30B-A3B and Qwen3-235B-A22B decreases, which take 72--83\,s across EP2--EP8 and 136--146\,s across EP4--8, respectively. This is because the DeepGEMM kernels greatly accelerate FP8 expert computation, compared with vLLM default Triton implementation on BF16. The \load time of \sys barely changes compared with BF16, as the graph topology is mostly the same, switching kernel functions. Yet, \sys still consistently achieves 96--97\% latency reduction, and completes in 2.8--2.9s on Qwen3-30B-A3B and 3.5s on Qwen3-235B-A22B.

\paragraph{Comparison with initialization without graphs.}
An alternative to accelerate cold start is to skip graph capture entirely and run in eager mode. The serving engine still needs to load kernels from \texttt{torch.compile} cache, but it is much faster than capturing hundreds of graphs. \autoref{fig:init-all-model} shows that vLLM without graphs achieves lower initialization latency (1.2--62\,s) but, as shown in \S\ref{sec:background}, it significantly degrades decoding performance, while \sys delivers the full performance of CUDA graphs (\autoref{fig:eval-tpot}). \sys achieves startup times comparable to or faster than eager mode, because all required binaries are already packed into a condensed \workspace with clear mappings to kernel functions. For example, Qwen3-14B DP8 starts in 5.0\,s without graphs versus 1.7\,s with \sys; Qwen3-235B-A22B EP8 starts in 62\,s without graphs versus 3.9\,s with \sys.

\subsubsection{Comparison with CUDA Checkpoint}
\label{sec:eval-cuda-ckpt}

\autoref{fig:cuda-ckpt} provides a detailed phase-level breakdown for four configurations, comparing vLLM, CUDA-checkpoint, and \sys.

CUDA-checkpoint eliminates warmup and graph capture by toggling back snapshotted GPU states. Using vLLM's \texttt{sleep} and \texttt{wakeup} mechanism, the vLLM model weights and KV cache are released before checkpointing to avoid storing the whole GPU memory (141GB per H200). 

Across different models, CUDA-checkpoint consistently restore serving engine in 5.7--6.1\,s, achieving 4.9--7.9x faster initialization compared with vLLM. However, \sys achieves 1.3--2.3\,s on these configurations, consistently outperforming CUDA-checkpoint by 2.6--4.4$\times$. The breakdown reveals that the toggle back is slower than the total time of loading binaries and graph in \sys. This is because \sys only saves the necessary CUDA states that are most time-consuming to capture, while creating others at runtime, whereas CUDA-checkpoint saves the entire CUDA state, incurring not only higher latency but also larger image size (\autoref{tab:hardware-cost}). 

\subsection{Serving Throughput Preservation}
\label{sec:eval-throughput}

Fast initialization is only useful if the reconstructed CUDA graphs produce the same serving performance as natively captured ones. As \sys targets the decoding phase of PD-disaggregated serving, we report TPOT to assess the throughput of the LLM serving. This value also reflects graph replay duration, which would be the same if the loaded graph is equivalent to the captured one. \autoref{fig:eval-tpot} compares mean TPOT between vLLM and \sys for representative configurations on both H200 and B200. We use the serving benchmark provided by vLLM to simulate incoming requests and test across 16--512 random prompts 10 times, with each prompt generating 128 tokens. 

\paragraph{No observable throughput degradation.}
Across all settings, the TPOT curves for \sys and vLLM with natively captured graphs overlap almost perfectly. The first subplot tests across DP1--4 on Qwen3-14B, confirming that \sys preserves serving performance for different data parallelism sizes. The second subplot further verifies that the performance holds when applying optimized custom kernels, such as DeepGEMM. Comparing the second and third subplots, we observe that for a large model that requires sharding to multiple GPUs, \sys still preserves same throughput as vLLM. And the fourth subplot confirms that this preservation generalizes to GPUs across different architectures. 

This result is expected by \sys{}'s library-agnostic design: its execution context materialization ensures that the restored engine contains the exact same kernel binaries, memory layout, regardless of what model is being served. This also confirms that the template-based reconstruction produces graph executables that are semantically equivalent to those produced by stream capture. The on-demand update to the template is very fast (\autoref{fig:graph-build-time}) and is shared by multiple engine steps due to continuous batching, resulting in no observable overhead. 

We also compare the tokens generated by \sys with vLLM and find that they are identical, confirming the correctness of CUDA graph and execution context restoration.

\subsection{Effectiveness of Templating}
\label{sec:eval-template-eff}

\subsubsection{Topology Sharing Across Models}
\label{sec:eval-topology}

\autoref{fig:template-percentage} reports the number of unique topologies (templates) identified by \sys for each model across 512 captured graphs. The number of templates ranges from 12 (Qwen3-235B-A22B FP8) to 25 (Qwen3-32B), yielding 95--98\% of graphs served via on-demand parameter update rather than full construction.

The fraction of templates is consistently low across model families we tested (Qwen3, Llama3 and Gemma3), indicating that the topological regularity exploited by \sys is a fundamental property of LLM inference, not specific to any architecture variant. The low template count means the graphs captured for different batch sizes execute the same sequence of kernel types with the same dependency structure, yet they can execute different kernels tuned for different matrix sizes, and with varied launch dimensions and arguments.

Notably, the template count does not grow with model size or complexity: the largest model, Qwen3-235B-A22B (FP8), requires only 12 templates, while the smaller Qwen3-32B requires 25. Besides, the template count is independent of parallelism degree for SPMD-style multi-GPU serving, as different ranks share the same execution flow as each works individually. The changes to matrix/vector size and rank identifier are covered by using different kernel handles, launch dimensions or arguments.

\subsubsection{Per-Graph Construction Cost}
\label{eval:on-demand}

\autoref{fig:graph-build-time} compares the average per-graph cost of three construction methods: native stream capture, template build via CUDA driver APIs, and on-demand parameter update of an existing template.

For Qwen3-14B DP4 and Qwen3-30B-A3B EP4, we observe that graph construction (31.1--69.5\,ms) is 1.9--2.9$\times$ faster than stream graph capture (59.7--198.6\,ms). However, because parallel graph construction results in driver contention (\autoref{app:parallel_graph_build}), sequentially building 512 graphs for Qwen3-30B-A3B still incurs $512 \times 69.5 \approx 35.6$\,s. This is not acceptable for elastic serving and it worsens for higher batch sizes. In-place graph update (0.98-2.89\,ms) further reduces the latency by 24--32$\times$ compared with graph construction, making it possible to execute on demand during serving. 

The on-demand update is fast because its required parameters have been organized during template construction (\S\ref{sec:design-grouping}), loading them and calling \texttt{cuGraphExecUpdate} to update node parameters in-place is much faster than re-instantiating a graph executable. As a result, the engine only needs to build very small number of graphs on initialization, effectively compresses initialization latency to a few seconds.

\subsubsection{Storage Cost Saving}
\label{sec:eval-cost}

\begin{table}[t]
\centering
\caption{Storage cost comparison between CUDA-checkpoint and \sys. L3, G3 and Q3 indicate Llama3, Gemma3 and Qwen3 respectively.}
\label{tab:hardware-cost}
    {
    \begin{tabular}{l|cccc}
        \toprule
        Model & L3-8B & G3-12B & Q3-14B &  Q3-235B \\
        Parallel & DP1 & DP1 & DP1 & EP8 \\ \hline
        Image size & 3.9\,GB & 6.6\,GB & 3.7\,GB & -- \\ \hline
        Archive size & 976\,MB & 1.3\,GB & 1.1\,GB & 2.2\,GB \\
        \bottomrule
    \end{tabular}
    }
\end{table}

\autoref{tab:hardware-cost} compares the storage cost between CUDA-checkpoint and \sys. For CUDA-checkpoint, model weights and KV cache are released before checkpointing.

\sys's \workspace is 4--5$\times$ smaller than CUDA-checkpoint's image. For example, on Qwen3-14B, the \workspace is 1.1\,GB versus 3.7\,GB for CUDA-checkpoint (3.4$\times$ smaller). This is because CUDA-checkpoint captures the entire process state, whereas \sys stores only graph metadata and kernel binaries, which are compact and rank-independent. Because \sys allows different ranks to share the same set of kernel binaries and graph metadata, even for Qwen3-235B-A22B EP8, the \workspace size is just 2.2\,GB. While CUDA-checkpoint doesn't even support expert parallelism, \sys creates its \workspace with one GPU.

\paragraph{Composition of \workspace.}
The \workspace consists of two parts: serialized graph metadata (topology, node parameters, and grouping manifest) and extracted kernel binaries. For Qwen3-235B-A22B EP8, the largest model evaluated, the total \workspace is 2.2\,GB, of which kernel binaries account for 1.4\,GB. The binary graph format (\S\ref{sec:impl-graph}) keeps the graph metadata compact, and \sys parses 512 serialized graphs in under 100\,ms.

%% file: related.tex
\section{Related Work}
\label{sec:related}

\subsection{LLM Cold Start Optimization}
A growing body of work has sought to reduce cold-start latency in elastic and serverless LLM serving. Many of these efforts focus on model loading. ServerlessLLM~\cite{fu2024serverlessllm} caches model weights on local SSDs, while BlitzServe~\cite{zhang2025blitzscale}, $\lambda$Scale~\cite{yu2025lambda}, and Tensor R-Fork~\cite{lmsys-rfork} leverage RDMA to fetch weights from peer instances. HydraServe~\cite{lou2026hydraserve} further overlaps model fetching with other stages of cold start, whereas WarmServe~\cite{lou2025warmserve} and InstaInfer~\cite{sui2024pre} proactively reduce startup latency by predicting future workloads and preparing resources in advance. To mitigate environment initialization overhead, DeepServe~\cite{hu2025deepserve} uses pre-warmed pods, following a broader line of work on container pre-warming~\cite{bhasi2021kraken,cadden2020seuss,fuerst2021faascache,lin2021flashcube,brooker2023demand,roy2022icebreaker}. Medusa, in contrast, targets CUDA graph capture overhead. However, unlike \sys, it materializes only graph topology and does not preserve the execution context required for general graph restoration.

\subsection{Parallelism Hot Switching in LLM Serving}
Recent work has begun to explore dynamic parallelism reconfiguration in LLM serving. LoongServe~\cite{wu2024loongserve} introduces elastic sequence parallelism to adapt the degree of parallelism to different requests; Gyges~\cite{chen2025gyges} and Flying Serving~\cite{gao2026flying} perform parallelism transformation to switch running instances across parallelism strategies as request context lengths vary. Elastic expert parallelism (EP)~\cite{mooncake2026elasticep} and Expert-as-a-Service (EaaS)~\cite{liu2025expert} have recently been proposed to enable fine-grained scaling and improve fault tolerance for serving MoE models. A common challenge across these systems is that changing the parallelism strategy also changes the computation logic, which requires recapturing CUDA graphs and thus incurs substantial reconfiguration overhead. 

\subsection{GPU Checkpoint/Restore}
GPU checkpoint/restore (C/R) approaches have been extensively studied. Existing system-level GPU C/R techniques can be broadly divided into driver-integrated and interception-based approaches. Driver-integrated C/R is vendor-specific; NVIDIA has recently added such support to its proprietary CUDA driver~\cite{nvidia2025cudacheckpoint,stoyanov2025criugpu}, but the current implementation still lacks support for IPC memory and therefore cannot be directly applied to multi-GPU distributed inference.  

On the other hand, interception-based systems~\cite{chaudhary2020balancing,eiling2022cricket,garg2018crum,jain2020crac,nukada2023efficient,takizawa2009checuda} record CUDA resource state during normal execution and reconstruct it during restore via CUDA driver API replay. Recent systems have substantially improved the efficiency of this line of work: PhOS~\cite{wei2025phoenixos} uses validated speculation to enable concurrent GPU C/R, while GCR~\cite{zeng2026gpu} uses control/data separation to reduce C/R latency and runtime overhead. However, because these systems restore CUDA process state through API replay, they do not directly support restoring captured CUDA graphs and are therefore complementary to \sys.

%% file: conclusion.tex
\section{Conclusion}
Modern LLM service providers increasingly rely on autoscaling to improve GPU resource efficiency, yet CUDA graph capture keeps LLM serving cold starts in the tens of seconds to minutes. We presented \sys, a template-based CUDA graph context materialization system that removes this bottleneck by persisting not only graph topology but also the execution context required for replay. By combining deterministic memory layouts, kernel binary materialization, topology-based templating, and single-GPU offline capture for multi-GPU distributed inference, \sys enables efficient, kernel-agnostic CUDA graph restoration without hand-crafted patching rules or heavyweight process-level checkpointing. Integrated with vLLM, \sys reduces cold-start latency by up to 99\%.

\clearpage

%% file: appendix.tex
\clearpage

\appendix

\section{Parallel CUDA Graph Construction}
\label{app:parallel_graph_build}

We build a standalone test that allocates different numbers of threads to build 80 graphs, each with 500 dummy nodes, in parallel. \autoref{tab:parallel-graph-build} shows that wall time is almost consistent when increasing number of threads, and per API call time increases, confirming the contention between CUDA Graph driver APIs.

\begin{table}[h]
\centering
\caption{Durations of building 80 graphs each with 500 dummy nodes using different number of threads.}
\label{tab:parallel-graph-build}
\begin{tabular}{c|ccc}
\toprule
\#Threads & Wall Time & AddNode & Instantiate \\ \hline
1         & 28.52 ms              & 0.70 $\mu$s                  & 0.8 $\mu$s                 \\ 
2         & 29.32 ms              & 1.44 $\mu$s                  & 1.6 $\mu$s                 \\
4         & 36.13 ms              & 3.52 $\mu$s                  & 3.8 $\mu$s                 \\
8         & 37.54 ms              & 7.31 $\mu$s                  & 6.6 $\mu$s             \\
\bottomrule
\end{tabular}
\end{table}

\clearpage
\section{Example Kernel Node Structure}
\label{app:kernel_node_example}

\begin{verbatim}

"id": 7,
"type": "KernelNode",
"params": {
    "blockDimX": 384,
    "blockDimY": 1,
    "blockDimZ": 1,
    "gridDimX": 2,
    "gridDimY": 62,
    "gridDimZ": 1,
    "sharedMemBytes": 206044,
    "kernel_node_attrs": {
        "attrQueryAvailable": true,
        "clusterDimX": 2,
        "clusterDimY": 1,
        "clusterDimZ": 1,
        "clusterSchedulingPolicyPreference": 1,
        "memSyncDomainMapDefault": 0,
        "memSyncDomainMapRemote": 1
    },
    "kernelParams": [
        {
            "index": 0,
            "offset": 0,
            "size": 1720
        }
    ],
    "extra": [
        "CU_LAUNCH_PARAM_BUFFER_SIZE",
        1720,
        "CU_LAUNCH_PARAM_BUFFER_POINTER",
        "null",
        "CU_LAUNCH_PARAM_END"
    ],
    "extra_argBuffer_hex": "3e0000003e00000003000000eeffffffe4388ee304000000f7ffffffe
    4388ee303000000fcffffff0000008001000000000102030408101820283037050911192129313806
    0a121a222a3239070b131b232b333a0c141c242c343b0d151d252d353c0e161e262e363d0f171f272
    f000000000000000000000000000000000000000.....1b600000 (3440 digits in total)",
    "function_name": "nvjet_tst_168x128_64x5_1x2_h_bz_TNN",
    "kernel_source_binary_hash": 6788486540864509700,
    "func_attrs": {
        "max_dynamic_shared_size_bytes": 206044,
        "preferred_shared_memory_carveout": -1,
        "cluster_scheduling_policy_preference": 0,
        "required_cluster_width": 0,
        "required_cluster_height": 0,
        "required_cluster_depth": 0
    }
}

\end{verbatim}

\clearpage